\documentclass[aps,preprint,letterpaper,amsmath,amssymb,floatfix]{revtex4}

\usepackage{graphicx}
\usepackage{dcolumn}
\usepackage{bm}

\newcommand{\gmol}{\ensuremath{\gamma_\mathrm{mol}}}

\begin{document}

\preprint{v8}

\title{Measuring the Molecular Polarizability of Air}

\author{M.J.~Madsen}
\author{D.R.~Brown}
\author{S.R.~Krutz}
\author{M.J.~Milliman}
\affiliation{Department of Physics, Wabash College, Crawfordsville, IN  47933}

\date{\today}

\begin{abstract}
We present an update of the ``refractive index of air'' experiment commonly used in optics and undergraduate advanced labs.  The refractive index of air is based on the average molecular polarizability, which we measured from the period of the phase shift in a Michelson interferometer as a function of pressure.  Our value of the average molecular polarizability of air is $\gmol=2.133\pm0.032\times10^{-29}$~m${}^3$~(95\% CI) and from this we find the refractive index of air at atmospheric pressure to be $n=1.0002651(66)$, which is in agreement with the accepted value of $n=1.000271375(6)$.

\end{abstract}

\maketitle

\section{Introduction}

Measuring the refractive index of air using a vacuum cell in one arm of a Michelson interferometer is a common experiment in optics and undergraduate advanced labs \cite{Michelson}.  In a Michelson interferometer an interference pattern is produced by splitting a beam of light into two paths, bouncing the beams back and recombining them. The different paths can be of different lengths or travel through different materials.  This creates alternating interference fringes that can be seen at the output of the interferometer.  Typically, the refractive index of air experiment consists of setting up an interferometer with a vacuum cell in one arm.  Students count the number of bright and dark fringes that change as air is pumped from the cell \cite{jeppesen,yap,hey}.  Based on this number, students calculate the index of refraction of the air in the cell at atmospheric pressure.  However, this measurement does not explain why the refractive index changes as a function of pressure nor does it measure an intrinsic property of the gas molecules.  Rather, the refractive index of air depends on the particular pressure, temperature, and humidity content of the room during the experiment, which is equivalent to measuring the phase of the interferometer for a given set of conditions.

We present an alternative approach to this experiment based on the average electric polarizability of a collection of molecules in response to an electromagnetic wave, an intrinsic property of the gas that depends on the electron cloud around the molecules.  We show how the polarizability can be found from the intensity versus pressure data in the output of a Michelson interferometer.  We updated the usual experiment apparatus by adding two sensors to the Michelson interferometer, an apertured light sensor on the output of the interferometer and a pressure sensor to the vacuum cell, in order to measure the output intensity as a function of pressure.  Finally, we show how the refractive index of air can then be calculated from the polarizability for a given pressure and temperature.

\section{Model}

\begin{figure}[tbp]   
\centering
\includegraphics[width=8cm,keepaspectratio]{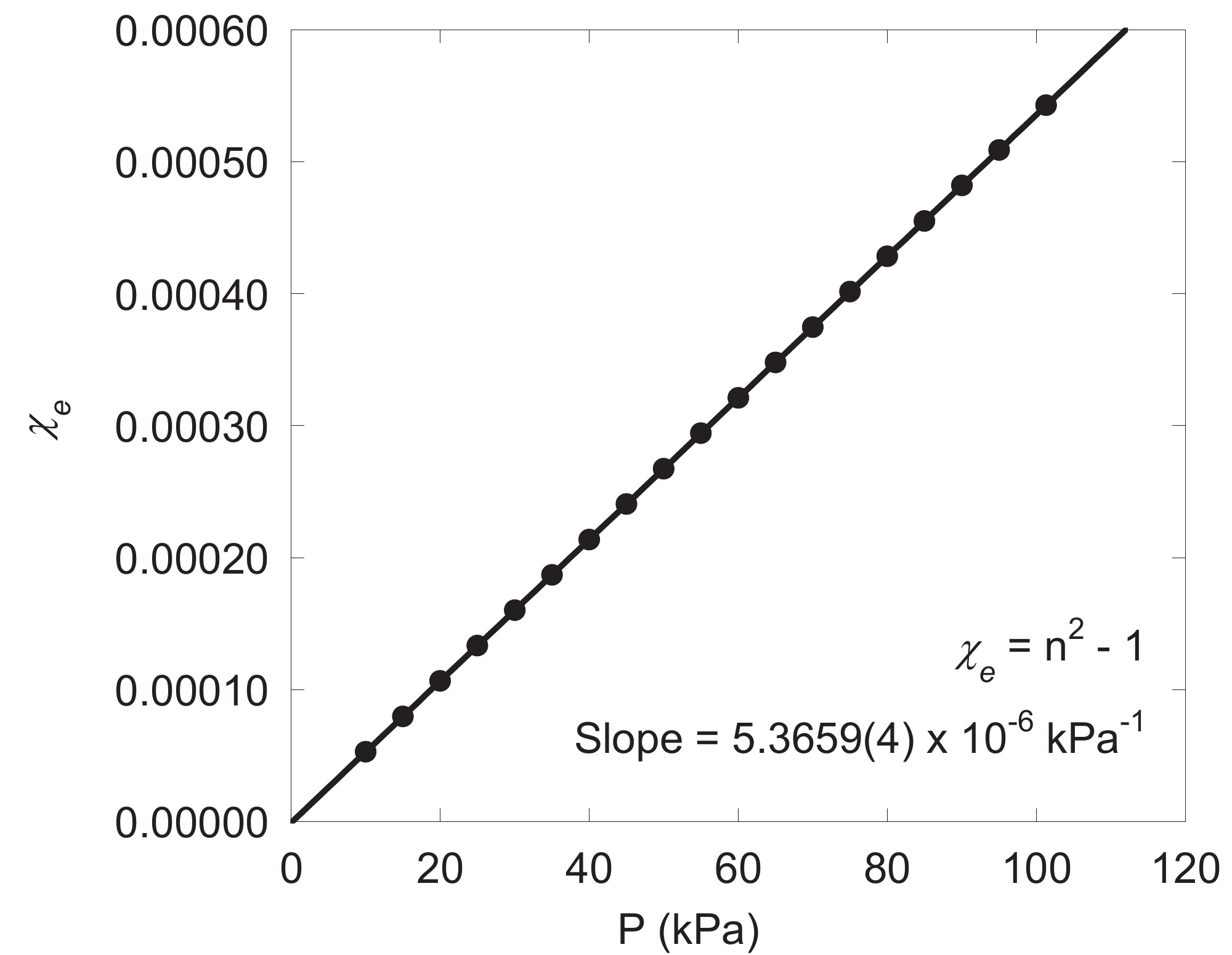}  
\caption{The dielectric susceptibility $\chi_e= n^2-1$ as a function of pressure $P$ for air starting at NTP, relative humidity of 50\%, and vacuum wavelength of $\lambda_0=632.991$~nm.  This linear interpolation is based on an emperical model for the refractive index of air \cite{Edlen,Ciddor}.  The interpolation was calculated using the ``Refractive Index of Air Calculator Based on Modified Edl\'en Equation'' \cite{TOOLBOX}. The slope of the $\chi_e$ vs. $P$ line is $5.3659(4)\times10^{-6}$~kPa${}^{-1}$ (95\% CI).}
\label{Edlen}
\end{figure}

The refractive index of air is an important parameter in a number of precision range-finding measurements and can be calculated precisely as a function of wavelength, pressure, temperature, and water content using empirical equations, known as the Edl\'en Equation \cite{Edlen,Ciddor}.  The National Institute of Standards and Technology (NIST) maintains an on-line calculator for precision data on the refractive index of air based on user-input parameters \cite{TOOLBOX}.  At a vacuum wavelength ($\lambda_0=632.991$~nm) for a Helium-Neon laser, NTP, and a relative humidity of 50\%, the refractive index of air (or equivalentely the dielectric susceptibility $\chi_e = n^2 -1$) is approximately linear for changes in pressure from atmosphere down to about 1~kPa, shown in Fig.~\ref{Edlen}.  We use this linear approximation to find a simple relationship between the air pressure in the vacuum cell in one arm of our Michelson interferometer and the phase difference (and thus intensity output) between the two optical waves in the two arms of the interferometer.

Atoms and molecules respond to external electric fields by creating an induced polarization.  In a mixed-element gas such as air, the polarization of the constituent molecules can be complex.  However, the average dipole moment of the medium $\langle \vec{p} \rangle$ can be measured as the net response of the gas to the effective electric field $\vec{E^\prime}$:
\begin{equation}
\langle \vec{p} \rangle = \epsilon_0 \gmol \vec{E^\prime},
\end{equation}
where $\epsilon_0$ is the permittivity of free space and the molecular polarizability, $\gamma_\mathrm{mol}$, depends on both the molecular structure and components of the gas \cite{Born and Wolf,Jackson}.  The macroscopic polarization vector $\vec{P} = \eta \langle \vec{p} \rangle$, where $\eta$ here is the number of molecules per unit volume (typically denoted as $n$, but since we use $n$ for the refractive index, we will indicate number density with $\eta$).  Since the macroscopic polarization is proportional to the dielectric susceptibility ($\chi_e$) of a substance, we find that the dielectric susceptibility is related to the molecular polarizability \cite{Jackson}:
\begin{equation}
\chi_e = \frac{\eta \gmol}{1-\frac{1}{3}\eta\gmol}.
\label{chie1}
\end{equation}
(For optical wavelengths, the dielectric susceptibility is related to the refractive index $n$, $\chi_e = n^2 - 1$.) We want an expression for the number density as a function of dielectric susceptibility, so we solve Eq.(\ref{chie1}) for $\eta$.  Since gases at NTP typically have a dielectric susceptibility on the order of $10^{-3}$, we keep only the first order term in $\chi_e$ which gives a number density that is approximately proportional to the dielectric constant,
\begin{equation}
\eta \approx \frac{\chi_e}{\gmol}.
\label{eta1}
\end{equation}
We model the air as an ideal gas so that for a gas pressure $P$ and temperature $T$, the number density $\eta$ is proportional to the pressure $\eta = P / kT$, where $k$ is Boltzmann's constant and $T\sim293$~K.  Using this and Eq.(\ref{eta1}), we find that the pressure is proportional to the dielectric susceptibility, $P \approx (kT/\gmol)\chi_e$, as was seen in Fig.~\ref{Edlen}.  From this we find that the predicted value, based on the Edl\'en equation calculator from NIST, of the molecular polarizability is $\gmol=2.1865(22)\times 10^{-29}$~m${}^3$.  Since the susceptibility for gases is small, we can find the refractive index of air $n$ as a function of pressure, keeping only the first order,
\begin{equation}
n\approx 1+ \frac{\gmol P}{2 k T}.
\label{n1}
\end{equation}

\begin{figure}[htbp]   
\centering
\includegraphics[width=8cm,keepaspectratio]{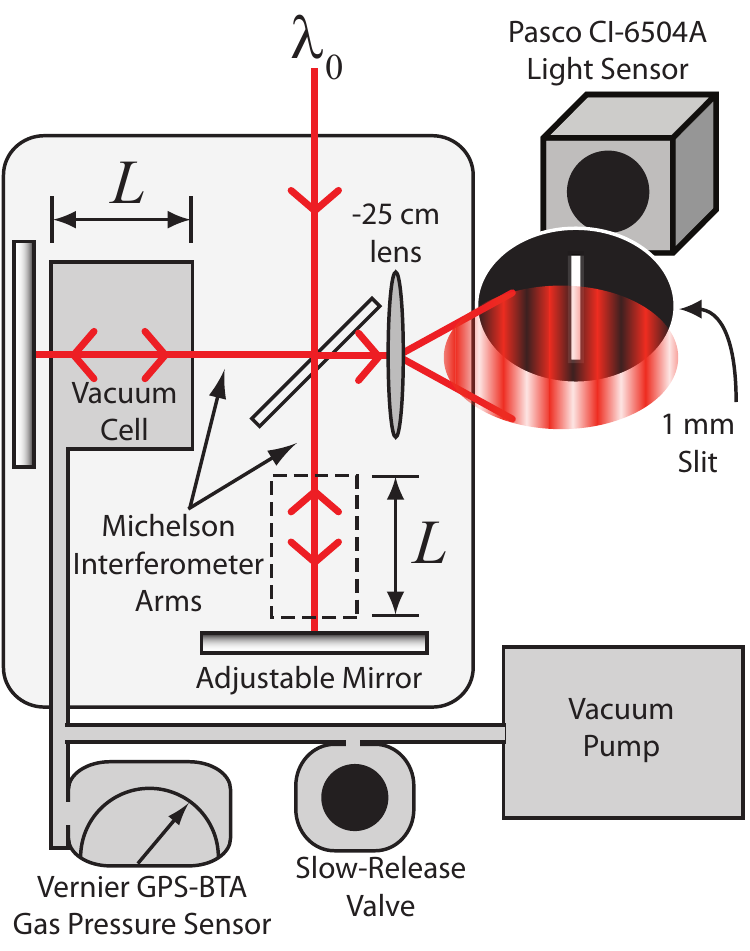}  
\caption{A Helium-Neon laser was coupled to a Michelson interferometer with a vacuum cell of effective length $L=0.02565\pm0.00037$~m inserted in one arm.  The alignment of one mirror was offset horizontally to give vertical output fringes that were about 10~mm wide.  A 1~mm slit was placed in front of our light sensor, giving us a fringe contrast of about 70:1.  The vacuum pump reduced the pressure in the cell to about 1~kPa, as measured with the gas pressure sensor.  A slow-release valve released the vacuum over a time period of about 200 seconds.}
\label{setup}
\end{figure}

We measured the change in the refractive index as a function of air pressure in a Michelson interferometer where the two paths have approximately equal physical lengths, as shown in Fig.~\ref{setup}.  The critical difference between the two arms of the interferometer can be characterized by two lengths: the optical path length of the vacuum cell $2n_\mathrm{cell}L$ in one arm and the reference optical path length $2n_\mathrm{ref}L$ in the other arm, noting that the light passes twice through both the vacuum cell and the reference length.  When the vacuum cell is filled with air $n_\mathrm{cell}=n_\mathrm{ref}=n_\mathrm{air}$ and path length difference ($\Delta\mathrm{OPL}$) between the two arms is zero.  However, if we let the index of refraction vary as a function of pressure as in Eq.(\ref{n1}), the optical path length difference becomes 
\begin{equation}
\Delta\mathrm{OPL} = 2L\left(1 + \frac{\gmol P}{2kT} - n_\mathrm{ref}\right).
\label{PLD}
\end{equation}
We are only interested in a phase difference that changes with pressure; the constant terms in Eq.(\ref{PLD}) give rise to a constant offset phase in the output of the interferometer.  The output intensity $I$ of the interferometer depends on this phase difference $\Delta\phi$ between the two arms.  As the phase difference increases with pressure, the intensity oscillates between a maximum, $I_0$, and zero intensity, $I=I_0 \cos^2(\Delta \phi/2)$.  Since the phase difference between the two arms depends on the optical path length difference, $\Delta\phi = 2\pi \Delta\mathrm{OPL}/\lambda_0$, the phase difference as a function of pressure is thus,
\begin{equation}
\Delta \phi= 2\pi \frac{L \gmol}{\lambda_0 kT}P + \Delta\phi_0
\end{equation}
where the phase offset $\Delta\phi_0$ depends on the reference arm described above and is constant and passively stable over long time scales. 

The output intensity of the interferometer is thus
\begin{equation}
I=I_0 \cos^2\left(2 \pi \frac{P}{\Pi} + \Delta\phi_0\right)
\label{Imodel}
\end{equation} 
where the period of the oscillations is 
\begin{equation}
\Pi=\frac{2 \lambda_0 kT}{L \gmol}.
\label{eqn:piA}
\end{equation}
Using this relationship we can find the molecular polarizability of air from the period of the intensity-versus-pressure data.  Once we have the molecular polarizability, we can use Eq.(\ref{n1}) to find the refractive index of air at NTP.

\begin{figure*}[tbp]   
\centering
\includegraphics[width=\textwidth,keepaspectratio]{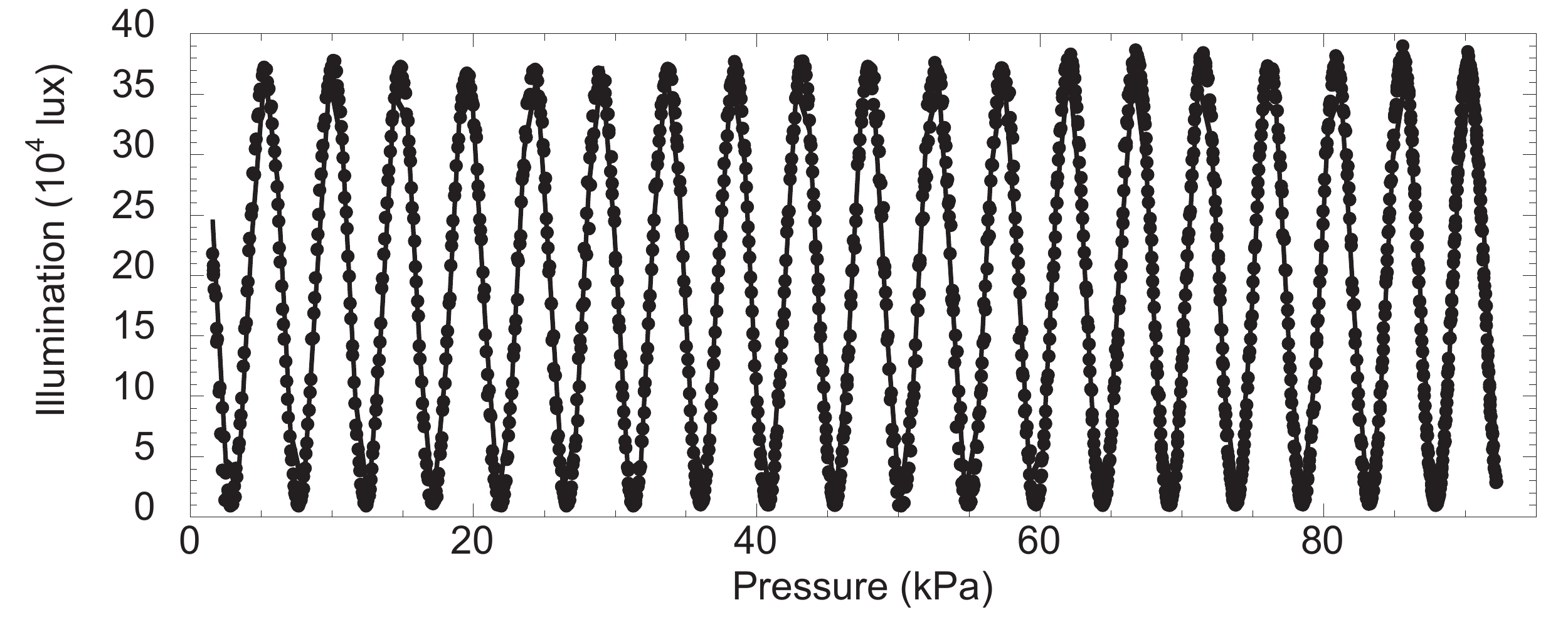} 
\caption{The output fringe intensity of our interferometer as a function of the pressure in the vacuum cell.  The sinusoidal dependence is due to the change in refractive index with pressure, and thus in the effective phase between the two arms of the interferometer.  The period of oscillation is $\Pi = 9.430 \pm 0.015$~kPa~(95\% CI) based on the statistical average of our four trials.}
\label{data}
\end{figure*}

\section{Methods}

We coupled a Helium-Neon laser (vacuum wavelength $\lambda_0 = 632.991$~nm) to a Michelson interferometer \cite{PMT} and aligned both arms so that the path-length difference was approximately equivalent, Fig.~\ref{setup}.  The beams were slightly mis-aligned by adjusting one mirror slightly off-center in the horizontal direction in order to produce a wide vertical fringe pattern in the output of the interferometer.  Any change in the phase difference makes the vertical fringe pattern move horizontally.  We magnified the fringes using a $-25$~cm focal length lens and detected the output intensity $I$ using a 1-mm-wide vertical slit in front of a photodetector \cite{PLS}.  A single fringe was approximately 10~mm wide, giving us a fringe contrast of $\sim$70:1 on the detector.

A vacuum cell in the second interferometer arm, shown in Fig.~\ref{setup}, had a total length of $3.2\pm0.1$~cm (95\% CI), with two windows of thickness $0.318\pm0.025$~cm (95\% CI), giving an effective length of the vacuum portion of the cell of $L=0.02565\pm0.00037$~m (95\% CI). The vacuum cell was aligned perpendicular to the beam in the second arm of the interferometer by using the back reflections from the glass faces of the cell.  A slight vertical offset in the exiting beam was maintained to ensure that the main output beam was not disrupted.

The pressure $P$ in the vacuum cell was measured using a gas pressure sensor \cite{GPS} and was pumped down to $\sim1$~kPa.  The vacuum was released slowly with a manual valve so that the total time for the pressure to reach atmospheric pressure from the initial pressure was about 200 seconds (with a data sampling rate of about 20 samples/second).  This prevented a systematic effect due to the vacuum cell windows moving with a faster pressure release.  Data from both the pressure sensor and the light sensor were collected simultaneously on a computer \cite{vernier}.  The data collected, shown in Fig.~\ref{data}, had a clear sinusoidal trend while the residuals where random.  

We fit the intensity $I$ versus pressure $P$ data using a four-parameter sinusoidal fit based on Eq.(\ref{Imodel}),
\begin{equation}
I = I_0\cos^2\left(\frac{2\pi P}{\Pi_\mathrm{fit}}+\phi_0\right)+I_\mathrm{offset}
\end{equation}
where the only parameter of interest is the period of the oscillation $\Pi_\mathrm{fit}$.  The experiment was carried out four times with similar conditions and we found a period of $\Pi_\mathrm{fit} = 9.430 \pm 0.015$~kPa (95\% CI).  During all four experiments the room was maintained at room temperature, $T=293\pm1$~K (95\% CI).  We found, using Eq.(\ref{eqn:piA}), the molecular polarizability of air to be $\gmol=2.133\pm0.032\times10^{-29}$~m${}^3$~(95\% CI).  This result is in slight disagreement (about 2\%) with the value from NIST, based on the Edl\'en data (Fig.~\ref{Edlen}) $\gamma_\mathrm{mol,NIST}=2.1866\pm0.0074\times10^{-29}$~m${}^3$.  However, we found the refractive index of air at atmospheric pressure to be $n=1.0002651(66)$ which is in agreement with the accepted value of $n=1.000271375(6)$~\cite{TOOLBOX}. 

We have shown that by measuring the output intensity of the Michelson interferometer as a function of pressure in a vacuum cell, the molecular polarizability can be accurately measured.  We have also shown explicitly how the observed intensity depends on the pressure in the vacuum cell.  This experiment highlights the techniques of making a differential measurement, and is in good agreement with previous experiments.  A few simple improvements to the experiment could be made by fabricating a custom vacuum cell with external windows in order to reduce the systematic uncertainty in the measurement of the molecular polarizability.  Additionally, various pure gases (such as nitrogen, argon, etc.) could be measured and the results compared to known values of the dielectric constant for those gases.  Finally, we made a measurement of the refractive index of air at atmospheric pressure, as in the traditional experiments, but calculated the result based on our more fundamental measurement.

\begin{acknowledgments}

I would like to thank D.E.~Krause for his help and suggestions.

\end{acknowledgments}

\end{document}